\documentclass[journal=jcisd8,manuscript=article]{achemso}

\usepackage[T1]{fontenc} 
\usepackage{chemformula} 
\usepackage{algorithm}
\usepackage{algpseudocode}
\usepackage{siunitx}
\usepackage{subcaption}
\usepackage{tikz}
\usepackage{glossaries}
\usepackage{booktabs,caption}
\usepackage[flushleft]{threeparttable}
\usepackage{multirow}
\usepackage{nameref}

\usepackage{todonotes}

\newcommand{\refbyname}[1]{``\nameref{#1}``}

\newcommand{\refenergyshort}{DFT}
\newcommand{\refenergy}{$\omega$B97X-D/6-311++G**}

\newcommand{\SVSIXGAFF}{$3.12 \pm 2.23$}
\newcommand{\SVSIXGAFFTWO}{$2.93 \pm 2.12$}
\newcommand{\SVSIXNNP}{$1.63 \pm 0.89$}

\newcommand{\SELLERSGAFF}{$1.62 \pm 1.37$}
\newcommand{\SELLERSGAFFTWO}{$1.62 \pm 1.36$}
\newcommand{\SELLERSNNP}{$0.55 \pm 0.38$}
\newcommand{\SELLERSPARAMETERIZE}{$0.61 \pm 0.36$}
\newcommand{\ZINCMAE}{$0.32 \pm 0.68$}

\newacronym{mm}{MM}{molecular mechanics}
\newacronym{md}{MD}{molecular dynamics}
\newacronym{ff}{FF}{force field}
\newacronym{qm}{QM}{quantum mechanics}
\newacronym{nn}{NN}{neural network}
\newacronym{nnp}{NNP}{neural network potential}
\newacronym{dft}{DFT}{density functional theory}
\newacronym{ml}{ML}{machine learning}
\newacronym{sbdd}{SBDD}{structure-based drug discovery}
\newacronym{aev}{AEV}{atomic environment vector}
\newacronym{mae}{MAE}{mean absolute error}
\newacronym{gpu}{GPU}{graphical processing unit}

\author{Raimondas Galvelis}
\affiliation[Acellera]{Acellera Labs, C/ Doctor Trueta 183, 08005 Barcelona, Spain}
\altaffiliation{Contributed equally to this work}
\author{Stefan Doerr}
\affiliation[Acellera]{Acellera Labs, C/ Doctor Trueta 183, 08005 Barcelona, Spain}
\alsoaffiliation[CSL]{Computational Science Laboratory, Universitat Pompeu Fabra, PRBB, C/ Doctor Aiguader 88, 08003 Barcelona, Spain}
\altaffiliation{Contributed equally to this work}
\author{Jo\~{a}o M. Damas}
\affiliation[Acellera]{Acellera Labs, C/ Doctor Trueta 183, 08005 Barcelona, Spain}
\author{Matt J. Harvey}
\affiliation[Acellera]{Acellera Labs, C/ Doctor Trueta 183, 08005 Barcelona, Spain}
\author{Gianni De Fabritiis}
\affiliation[Acellera]{Acellera Labs, C/ Doctor Trueta 183, 08005 Barcelona, Spain}
\alsoaffiliation[CSL]{Computational Science Laboratory, Universitat Pompeu Fabra, PRBB, C/ Doctor Aiguader 88, 08003 Barcelona, Spain}
\alsoaffiliation[ICREA]{Instituci\'{o} Catalana de Recerca i Estudis Avan\c{c}ats (ICREA), Passeig Lluis Companys 23, 08010 Barcelona, Spain}
\email{gianni.defabritiis@upf.edu}

\title{A Scalable Molecular Force Field Parameterization Method Based on Density Functional Theory and Quantum-Level Machine Learning}

\begin{document}


\begin{tocentry}
\centering
\includegraphics{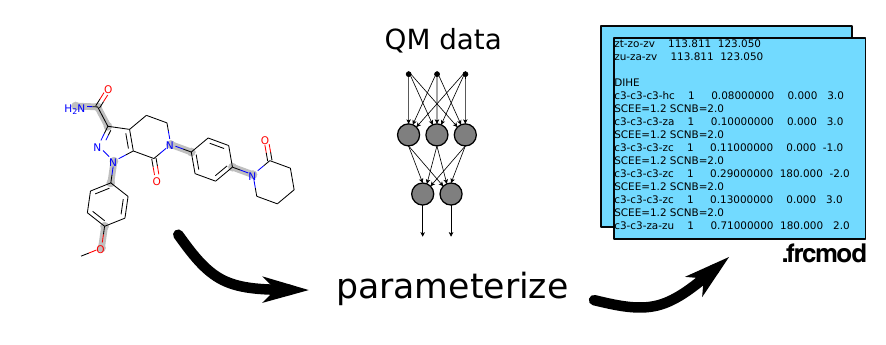}
\end{tocentry}

\begin{abstract}
Fast and accurate molecular \gls{ff} parameterization is still an unsolved problem. Accurate \glspl{ff} are not generally available for all molecules, like novel druglike molecules. While methods based on \gls{qm} exist to parameterize them with better accuracy, they are computationally expensive and slow, which limits applicability to a small number of molecules.
Here, we present an automated \gls{ff} parameterization method which can utilize either DFT calculations or approximate \gls{qm} energies produced by different \glspl{nnp}, to obtain improved parameters for molecules. We demonstrate that for the case of torchani-ANI-1x \gls{nnp}, we can parameterize small molecules in a fraction of time compared with an equivalent parameterization using DFT \gls{qm} calculations while producing more accurate parameters than  \gls{ff} (GAFF2).
We expect our method to be of critical importance in computational \glsentrylong{sbdd}. The current version  is available at \emph{PlayMolecule} (\url{www.playmolecule.org}) and implemented in HTMD, allowing to parameterize molecules with different \gls{qm} and \gls{nnp} options.
\end{abstract}

\section{Introduction}
\glsresetall 


In \gls{mm},  molecular interactions are represented by empirical potentials and their parameter sets. These parameter sets, called \glspl{ff}, are crucial for \gls{mm}'s accuracy and applicability. \gls{mm} has been successfully applied in large-scale biomolecular simulations in many cases ranging from  protein folding~\cite{LindorffLarsen2011}, protein-protein interactions~\cite{Plattner2017} to protein-ligand binding~\cite{buch2011}.
Typically, the development of a \gls{ff} is cumulative and collective effort focused on a particular subset of the chemical space. For example, the most popular biomolecular \glspl{ff} families AMBER~\cite{cornell1995second, maier2015ff14sb} and  CHARMM~\cite{mackerell1998all, huang2013charmm36} have parameters for proteins, lipids, DNA, etc. If one needs parameters for a particular molecule outside that chemical space, it has to be parameterized, which is a non-trivial, time-consuming and computationally expensive process.


In this work, we focus on small biologically-active (i.e. drug-like) molecules with $\sim$100 atoms. The accessible chemical space is estimated to span from  $10^{14}$ to $10^{180}$ molecules~\cite{Polishchuk2013}. Accessing accurate \gls{ff} parameters for any of these  molecules in a fast manner is critical for many fields of computational chemistry, especially computational \gls{sbdd}.

\begin{figure}
    \begin{subfigure}{0.45\textwidth}
        \includegraphics[width=\textwidth]{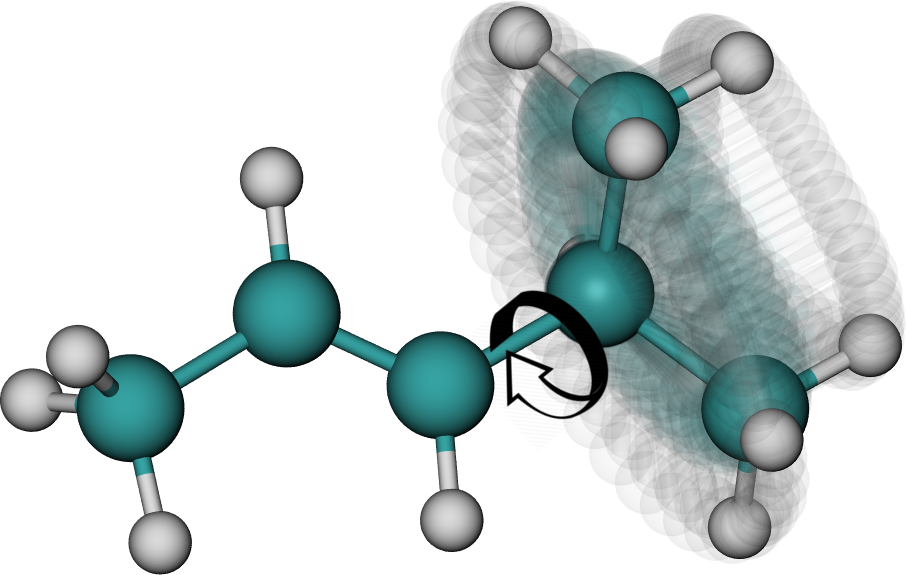}
        \caption{}
        \label{fig:demo_molecule}
    \end{subfigure}
    \begin{subfigure}{0.45\textwidth}
        \includegraphics[width=\textwidth]{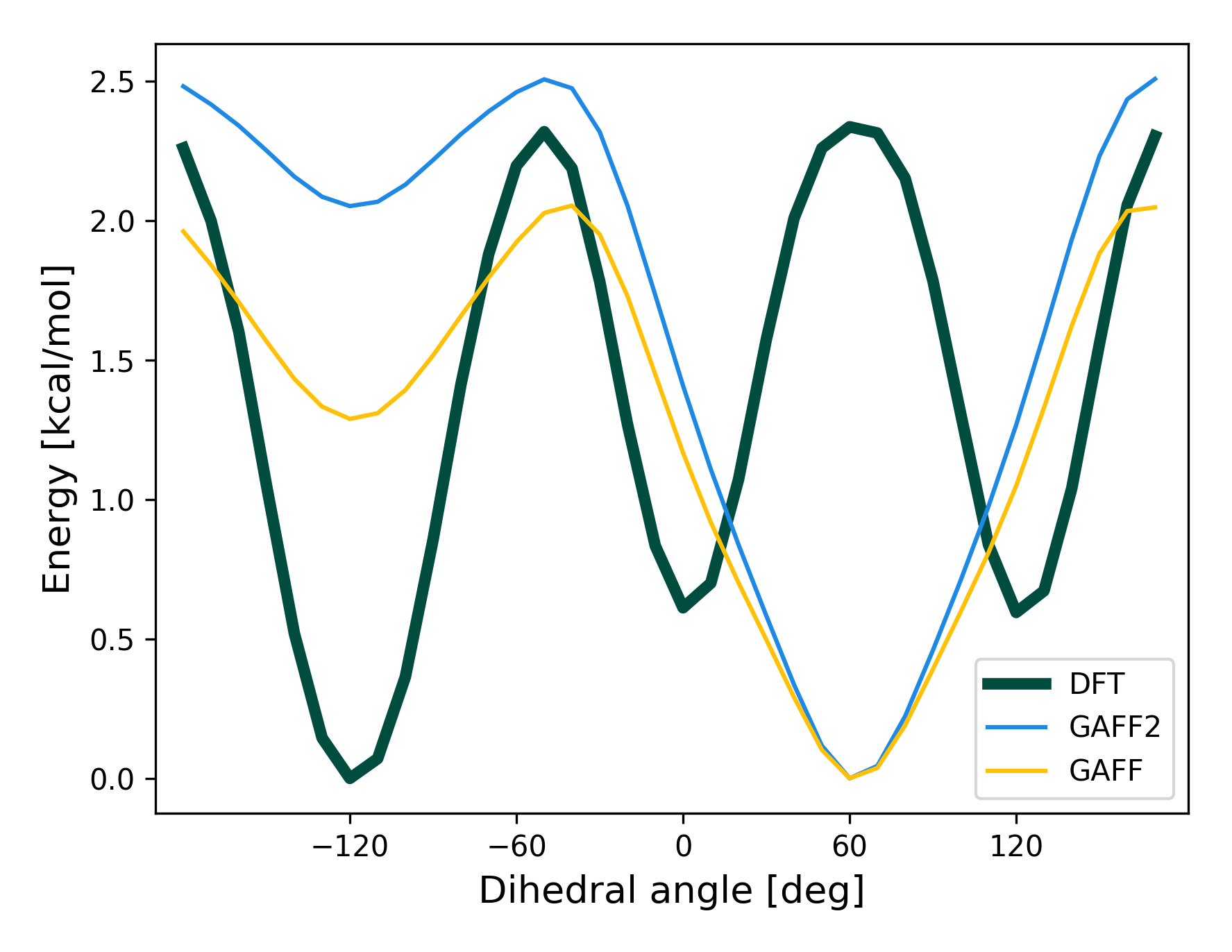}
        \caption{}
    \end{subfigure}
\caption{(a) 2-methyl-3-pentene is rotated around the marked bond and (b) the energies of the rotamers are calculated using \refenergyshort~(see section~\refbyname{sec:data} for details), GAFF and GAFF2 parameters. This highlights the limitation of the GAFF methods.}
\label{fig:gaff2fail}
\end{figure}

For both AMBER and CHARMM there are general \glspl{ff} (e.g. GAFF~\cite{Wang2004}, CGenFF~\cite{vanommeslaeghe2010charmm}), which extend the base \glspl{ff} with more chemical groups and intend to cover more molecules through heuristic pattern matching. Unfortunately, this approach does not always provide accurate \gls{ff} parameters, especially for dihedral angle parameters. In Figure~\ref{fig:gaff2fail}, we highlight this problem in a molecule, where an energy profile with GAFF parameters is in disagreement with the reference \gls{qm} calculations. For example, such inaccurate parameterization will result in poor computational \gls{sbdd} results, potentially failing to identify the best drug candidates.

Several methods (GAAMP~\cite{Huang2013}, ffTK~\cite{Mayne2013}, and ATB~\cite{Malde2011}) have been developed to circumvent such problems. Typically, they take one of these general \glspl{ff} as a base and improve the parameters of each individual molecule (mostly dihedral angle parameters, but also atomic charges) by fitting to \gls{qm} results. However, due to the high computational cost of \gls{qm}, this approach is only tractable for a limited number of small molecules.

Recent advances in \gls{ml} have lead to the development of \glspl{nnp}\cite{Behler_2017, Behler_2015}, which are trained on \gls{qm} data to predict the total energies of molecules at a fraction of the computational cost\cite{Behler_copper, smith_ani-1_2017, deep_potential, tensormol, ani1xcc, smith_less_2018, Hellstrom2018, torchani, SchNet, aimnet}. For example, ANI-1x was trained on organic molecules and achieved the \gls{mae} of 1.6~kcal/mol for the energies of COMP6 benchmark\cite{ani1xcc, smith_less_2018}, while it took less than a second per molecule. Such speed and accuracy can be exploited for \gls{ff} parameterization.


In this work, we develop a method, called \emph{Parameterize}, for the \gls{ff} parameterization of individual molecules. It combines a general \gls{ff} and \gls{nnp}: the initial \gls{ff} parameters are obtained with GAFF2, then selected dihedral angles are scanned and their parameters are fitted to ANI-1x\cite{ani1xcc, smith_less_2018} energies. This allows to improve the dihedral angle parameters, while keeping the computational cost low. The quality of \gls{ff} dihedral angle parameters is important for correct molecular shape, conformation distribution, and thermodynamic properties\cite{li2009silico}. We believe that the accuracy and speed of this method makes it practical for many applications, in particular computational \gls{sbdd}. \emph{Parameterize} is implemented with HTMD\cite{htmd} and available as an application on \emph{PlayMolecule} (\url{www.playmolecule.org}).\cite{restriction}

\section{Methods}

\begin{figure}
    \centering
    \resizebox{0.7\textwidth}{!}{\begin{tikzpicture}

\usetikzlibrary{arrows.meta}
\tikzset{%
node distance=2cm,
box/.style = {rounded corners, minimum width=4cm, minimum height=1.25cm,
              draw=orange, fill=orange!25, align=center, thick},
arrow/.style = {->, rounded corners, thick, >={Latex[width=2mm,length=3mm]}}
}

\node(mol)  [box, draw=blue, fill=blue!25] {Molecule};

\node (at)  [box, below of=mol, xshift=-3cm]   {Atom type \\ assignment};
\node (gas) [box, below of=at, xshift=-4.75cm] {Gasteiger charge \\ assignment};
\node (ip)  [box, below of=at]                 {Initial parameter \\ assignment};
\node (min) [box, below of=ip]                 {Minimization};
\node (bcc) [box, below of=min]                {AM1-BCC charge \\ assignment};

\node (dd)  [box, below of=mol, xshift=3cm] {Dihedral angle \\ selection};\node (rg)  [box, below of=dd]              {Rotamer \\ generation};
\node (rm)  [box, below of=rg]              {Constrained \\ minimization};
\node (ref) [box, below of=rm]              {Reference energy \\ calculation};

\node (nt)  [box, below of=bcc, xshift=3cm]  {New atom type \\ creation};
\node (fit) [box, below of=nt]               {Dihedral parameter \\ fitting};
\node (ff)  [box, below of=fit, draw=red, fill=red!25] {Force field};

\draw[arrow] (mol) -| (at);
\draw[arrow] (mol) -| (dd);

\draw[arrow] (at) -| (gas);
\draw[arrow] (at) -- (ip);
\draw[arrow] (gas) |- (min);
\draw[arrow] (ip) -- (min);
\draw[arrow] (min) -- (bcc);
\draw[arrow] (bcc) |- (fit);

\draw[arrow] (dd) -- (rg);
\draw[arrow] (rg) -- (rm);
\draw[arrow] (rm) -- (ref);
\draw[arrow] (ref) |- (fit);

\draw[arrow] (nt) -- (fit);
\draw[arrow] (fit) -- (ff);

\draw[arrow] (at) -| ([xshift=0.25cm]nt.north);
\draw[arrow] (ip) -| ([xshift=-0.25cm]nt.north);
\draw[arrow] (min.east) -- (rg.west);
\draw[arrow] (bcc.east) -- (rm.west);

\end{tikzpicture}}
    \caption{\emph{Parameterize} procedure: the orange boxes represent the parameterization steps, while the arrows show their dependencies (data flow). See the text for more details.}
    \label{fig:param_graph}
\end{figure}
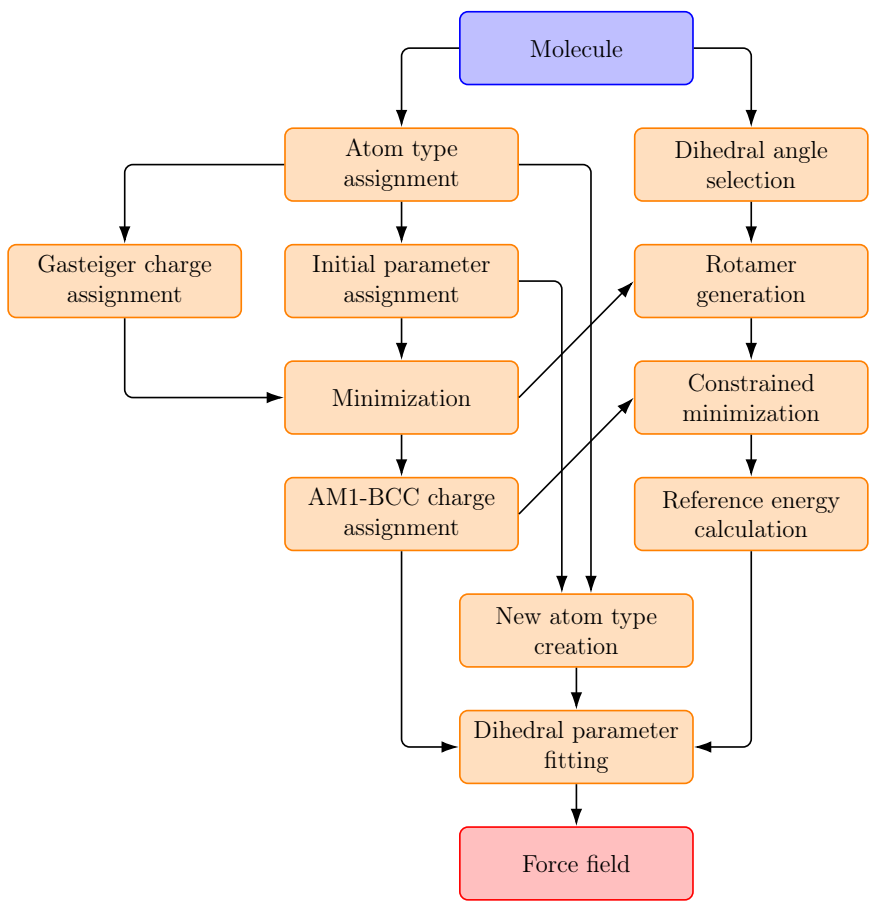

The parameterization method (Figure~\ref{fig:param_graph}) consists of three main parts. First, the initial \gls{ff} is constructed using GAFF2\cite{Wang2004} parameters and AM1-BCC\cite{bcc_charges_1, bcc_charges_2} atomic charges. Second, parameterizable dihedral angles (i.e. rotatable bonds) are selected and scanned by generating a set of rotamers. The rotamers are minimized with the initial \gls{ff} and their reference energies are evaluated with ANI-1x\cite{ani1xcc, smith_less_2018}. Finally, the dihedral angle parameters are fitted to reproduce the reference energies, resulting in an improved \gls{ff} for a target molecule. The following sections provide detailed descriptions of each part.

\subsection{Initial \gls{ff} parameters and atomic charges}\label{sec:initialparams}


GAFF\cite{Wang2004} is a general AMBER force field for drug-like molecules based on heuristic pattern-matching. We use its second version (GAFF2), as implemented in \emph{Antechamber}~\cite{antechamber}, to assign atom types and initial \gls{ff} parameters (bonded and Van der Waals terms) for the target molecule (Figure~\ref{fig:param_graph}). The use of GAFF2 ensures that our parameters are compatible with AMBER~\cite{cornell1995second, maier2015ff14sb} \gls{ff}  and can be used for bio-molecular simulations.


GAFF2 parameters are designed to be used with either with AM1-BCC\cite{bcc_charges_1, bcc_charges_2} or RESP\cite{esp_charges, resp_charges} charges. We use AM1-BCC, as it is computationally cheaper than RESP, which requires \gls{qm} calculations. Another problem with AM1-BCC (and RESP) is the charge dependency on the conformation of the target molecule, so the equilibrium conformation has to be used.

Also, we found out that geometry minimization of drug-like molecules with \gls{qm} occasionally results in topology and local geometry changes incompatible with the initial \gls{ff}. For example, a hydrogen atom moves to a different protonation state or an amine group minimizes to the pyramidal shape, while GAFF2 parameters keeps it planar.

We circumvent this issue by using intermediate Gasteiger charges\cite{gasteiger_charges}, which depend on molecular topology, but not conformation. First, the target molecule is minimized with the initial GAFF2 parameters and Gasteiger charges. Then, the minimized conformation of the molecule is used to compute AM1-BCC (Figure~\ref{fig:param_graph}). This scheme has a double benefit: we avoid computationally expensive minimization with \gls{qm} and ensure that our \gls{ff} can model the minimized molecule (i.e. the topology and local geometries are consistent).

\subsection{Dihedral angle selection and scanning}\label{sec:diheds}

\begin{figure}
\centering
\includegraphics[width=0.55\textwidth]{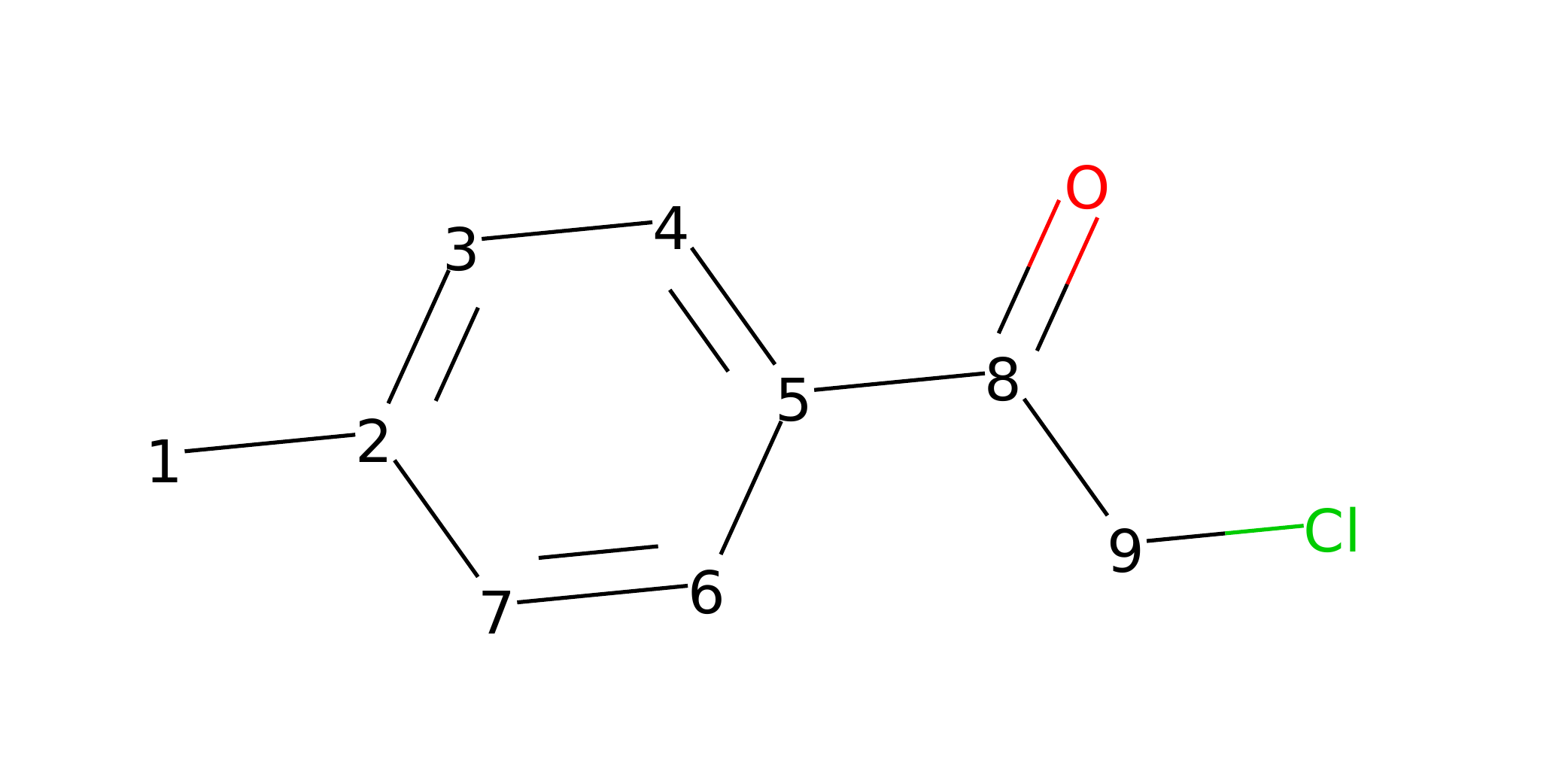}
\caption{An example molecule with two parameterizable dihedral angles: C4--C5--C8--C9 and C5--C8--C9-Cl. See section \refbyname{sec:diheds} for details.}
\label{fig:dihed_mol}
\end{figure}

Our selection of the dihedral angles for parameterization is based on a four-step scheme:
\begin{enumerate}
\item Select all covalent bonds between non-terminal atoms, except the bonds in rings. For symmetric molecules, only one of the equivalent bonds are chosen arbitrarily. In an example (Figure~\ref{fig:dihed_mol}), C1--C2, C5--C8, and C8--C9 bonds are selected. 
\item Exclude the bonds in the methyl groups, i.e. the C1--C2 bond is excluded.
\item Select the dihedral angles which contain the selected bonds their centers. For the C5--C8 bond, it results in four dihedrals: C4--C5--C8--O, C4--C5--C8--C9, C6--C5--C8--O, and C6--C5--C8--C9.
\item Select the dihedral angles which follow the longest chain, i.e. C4--C5--C8--C9 and C6--C5--C8--C9 have precedence over C4--C5--C8--O and C6--C5--C8--O (C4 and C6 are equivalent, so one of these dihedrals are chosen arbitrary). Additionally, priority is given to the dihedrals with the heaviest terminals, i.e. C5--C8--C9--Cl has precedence over C5--C8--C9--H.
\end{enumerate}

Each selected dihedral angle is scanned by generating 36 rotamers (every 10\textdegree) from the minimized conformation of the target molecule. The potential clashes between atom are removed with constrained minimization, where the selected dihedral angle is constrained and the remaining degrees of freedom are minimized with \gls{mm} using the initial GAFF2 parameters and AM1-BCC charges.

\subsection{Dihedral angle parameter fitting}


The reference energies of the minimized rotamers are computed with a \gls{nnp} (see section \refbyname{sec:torchani} for details). The high-energy rotamers (>\SI{20}{kcal/mol} above the minimum energy) are discarded to avoid non-physical conformations.


New atom types are created for atoms, which belong to the selected dihdrals angles. This ensures that our improved \gls{ff} does not conflict with the original \gls{ff}. The new atom types have the same charges, bonded and non-bonded parameters, except the dihedral angles parameters.


In AMBER~\cite{cornell1995second, maier2015ff14sb} \gls{ff}, the dihedral angle potential $E_\mathrm{dihed}(\phi)$ is expressed as a sum of six Fourier terms:
\begin{equation}
E_\mathrm{dihed}(\phi) = \sum^6_{n=1} k_n (1 + \cos(n \phi + \phi_n)),
\end{equation}
where $\phi$ is a dihedral angle value, $k_n$ is a force constant, and $\phi_n$ is a phase angle. For each parameterizable dihedral, a separate set of 12 parameters (6 force constants and 6 phase angles) is used.
The potential energy of the molecule consists of the parameterizable dihedral angle terms and other force field terms:
\begin{equation}
E_\mathrm{MM} = E_\mathrm{other} + \sum^{N_\mathrm{dihed}}_{i=1} E_{\mathrm{dihed,i}},
\end{equation}
where $N_\mathrm{dihed}$ is the number of parameterizable dihedral angles, including their equivalents. $E_\mathrm{other}$ includes all other force field terms (i.e. bonded terms, planar angle terms, non-parameterizable dihedral angle terms, and non-bonded terms), which are precomputed and kept constant during parameterization.

The fitting of dihedral angle parameters is performed by minimizing the RMSD of the reference energies and estimated MM energies:
\begin{equation}
\mathcal{L}_\mathrm{RMSD} = \sqrt{\frac{\sum^{N_\mathrm{rot}}_{i=1} (E_{\mathrm{ref},i} - E_{\mathrm{MM}, i} + C)^2}{N_\mathrm{rot}}},
\end{equation}
where $N_\mathrm{rot}$ is the number of rotamers and $C$ is an offset constant accounting for the absolute difference between the reference and \gls{mm} energies.
The $\mathcal{L}_\mathrm{RMSD}$ is a multi-modal function with respect to $k_n$ and $\phi_n$, which requires a global optimization.

For the global optimization, we use an iterative algorithm, where each iteration consists of two stages. In the first stage, the parameters of each dihedal angle are optimized individually with the na\"ive random search algorithm. The initial parameters are drawn from the random uniform distribution ($k_n \in [0, 10]$~kcal/mol and $\phi_n \in [0, 360)$\textdegree) followed by local minimization with L-BFGS. This assumes that the parameters of different dihedral angles are weakly correlated. In the second stage, we account for that correlation by optimizing all the dihedral angle parameters simultaneously with L-BFGS.

\subsection{Neural network potentials}\label{sec:torchani}

\Glspl{nnp} are classical models, which represent the energy potential of molecules with \glspl{nn}. \Glspl{nn} are universal function approximators, i.e. given enough  parameters a \gls{nn} can interpolate any sufficiently regular function \cite{HORNIK1989359}. In the case of \glspl{nnp}, \glspl{nn} are used to predict atomic energies from atomic environments.

\begin{figure}[t]
\centering
\includegraphics[width=0.3\textwidth]{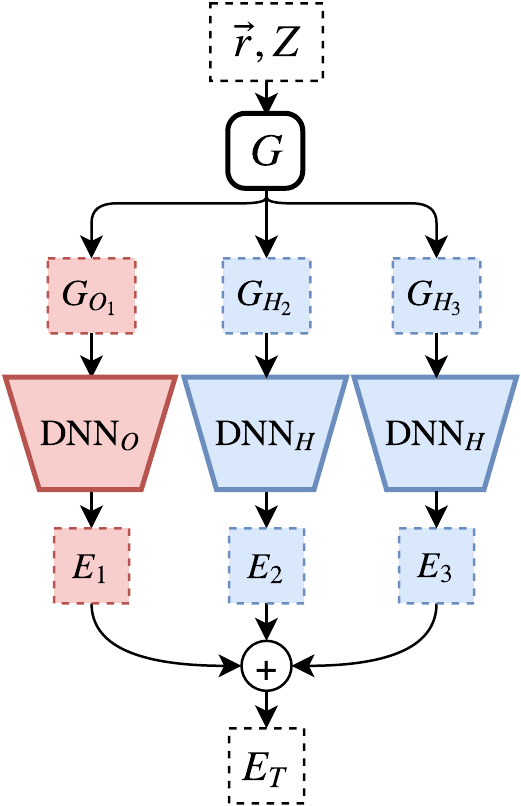}
\caption{An example of applying a \gls{nnp} to a water molecule, with $\vec{r}$ being the atom coordinates, $Z$ the elements, $G$ the featurization function producing the $G_{n}$ AEV of atom $n$, DNN the deep neural networks, $E_{n}$ the energy of atom $n$ and $E_{T}$ the total potential energy. As can be seen, the same hydrogen network is used to calculate the energies of both hydrogen atoms while the energy of the oxygen is calculated by a different oxygen network.}
\label{fig:nnp}
\end{figure}

TorchANI\cite{torchani} is an implementation of ANI-1x\cite{ani1xcc, smith_less_2018} with PyTorch\cite{pytorch}. ANI-1x computes the total energy of a molecular configuration as the sum of atomic energies (Figure~\ref{fig:nnp}). The atomic energies are computed with atomic \glspl{nn}. Each atomic \gls{nn} is a fully-connected \gls{nn} with three hidden layers using CELU activation functions and is trained to compute energies for a specific element. Currently, ANI-1x supports four elements (H, C, N, and O). The input to the atomic \glspl{nn} are the \glspl{aev}, which describe the local environment of each atom.


\Glspl{aev} are computed with the modified Behler-Parrinello (BP) symmetry functions\cite{smith_ani-1_2017}. The BP functions are atom-centred and translation/rotation invariant. They are truncated at a fixed radius to improve the scaling with the number of atoms and generalize over systems of arbitrary size.


ANI-1x consist of an ensemble of 8 \glspl{nnp} with different number of hidden layers and trained on different subsets of the training data\cite{ani1xcc, smith_less_2018}. The total energy is computed as a mean of the ensemble. The disagreements in the ensemble can be used an estimate of prediction error. For brevity, we refer to it as \gls{nnp}.

\subsection{Input and output}\label{sec:io}

The input to \emph{Parameterize} can be in Tripos MOL2 or MDL SDF molecular structures formats, which contain elements, topology, initial conformation, and molecular charge. Alternatively, the molecule can be entered as a SMILE string or with \emph{Kekule.js}\cite{kekule.js}, an integrated graphical chemical structure drawer. In case, the initial conformation is not preset, it is generated with \emph{RDKit}\cite{rdkit}.

The output is Tripos MOL2 and AMBER force field parameter modification format (FRCMOD) files. The former contains the atom types and the latter contains the corresponding \gls{ff} parameters. The files could be used with \emph{Antechamber}~\cite{antechamber} or HTMD~\cite{htmd} to build simulation systems. Note, if more than one parameterized molecule is used in the same system, the atom type conflicts have to be resolved manually.

\subsection{Evaluation datasets}\label{sec:data}


The accuracy of \gls{nnp} and our parameterization method is compared with \gls{qm}. The \gls{qm} calculations were performed at the \gls{dft} level of theory using $\omega$B97X-D \cite{wB97X-D} exchange-correlation functional and 6-311++G** basis set (i.e. \refenergy) with Psi4\cite{psi4}. For brevity, we refer to it as \refenergyshort.


For evaluation, we use three datasets. The first dataset consists of 45 molecules from \citet{Sellers} (17 molecules were skipped because they contained elements not supported by \gls{nnp}). The dataset contains drug-like fragments with various rotatable bonds.

\begin{figure}[t]
\centering
\begin{subfigure}[b]{0.41\textwidth}
\includegraphics[width=\textwidth]{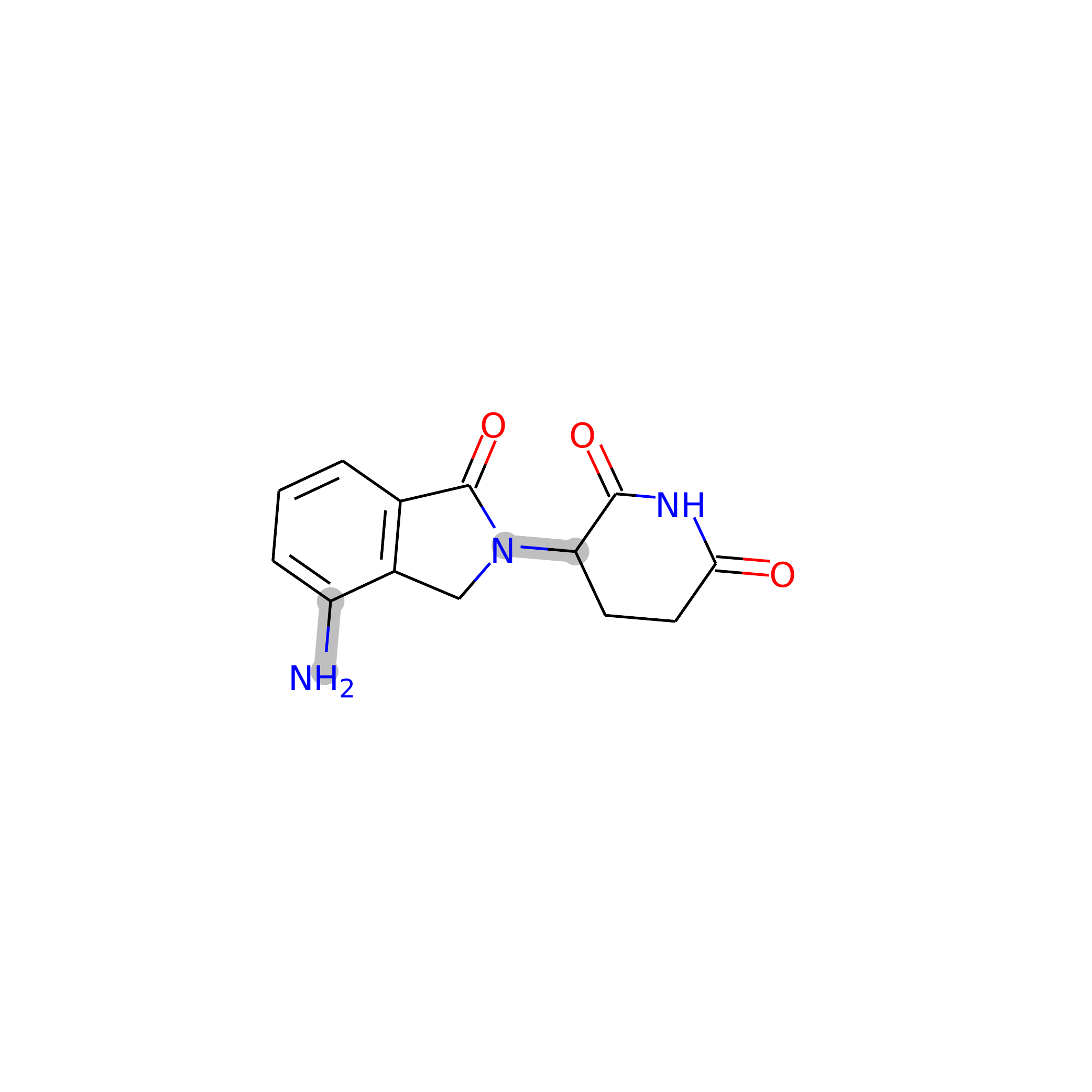}
\caption{Lenalidomide (LVY)}
\end{subfigure}
\begin{subfigure}[b]{0.41\textwidth}
\includegraphics[width=\textwidth]{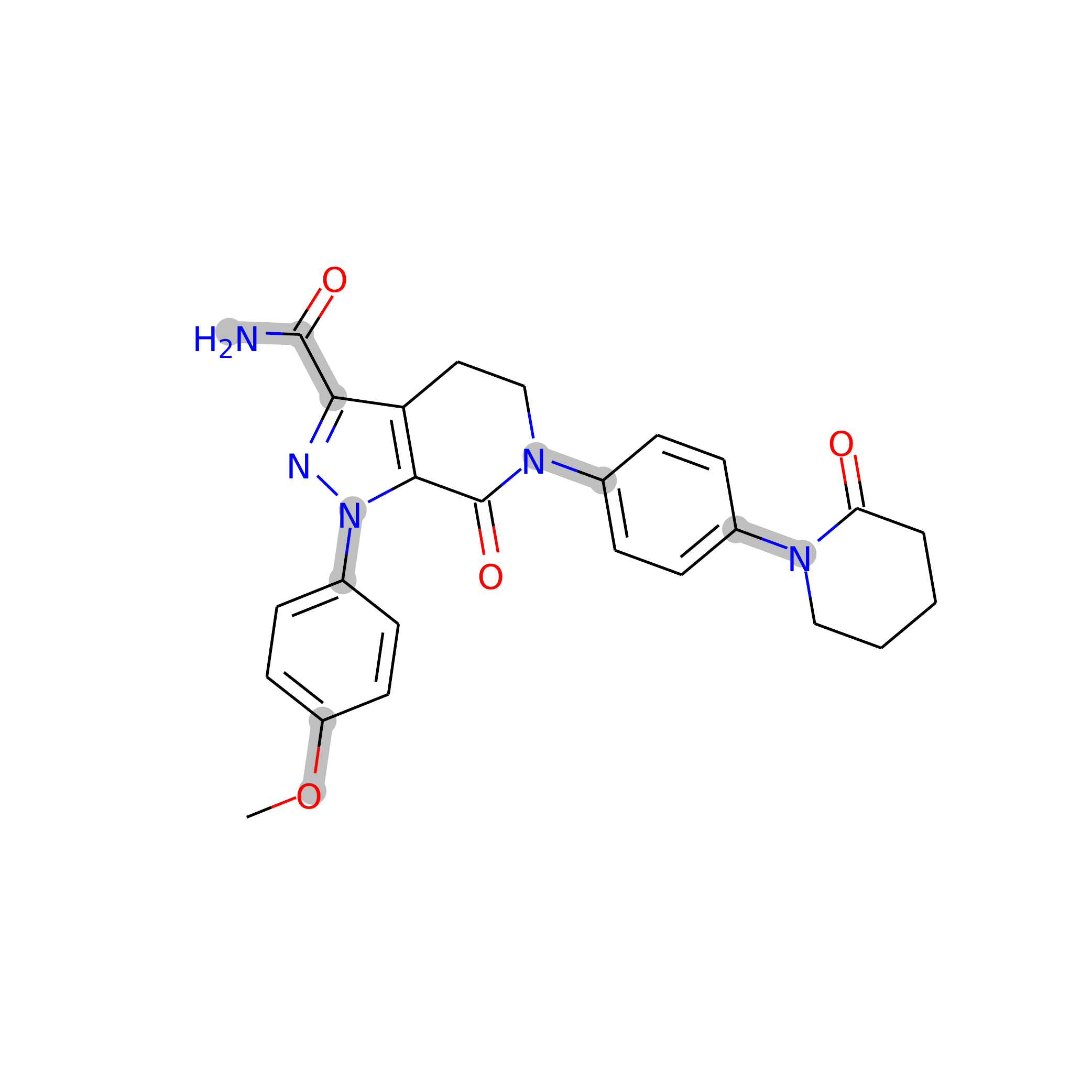}
\caption{Apixaban (GG2)}
\end{subfigure}
\begin{subfigure}[b]{0.41\textwidth}
\includegraphics[width=\textwidth]{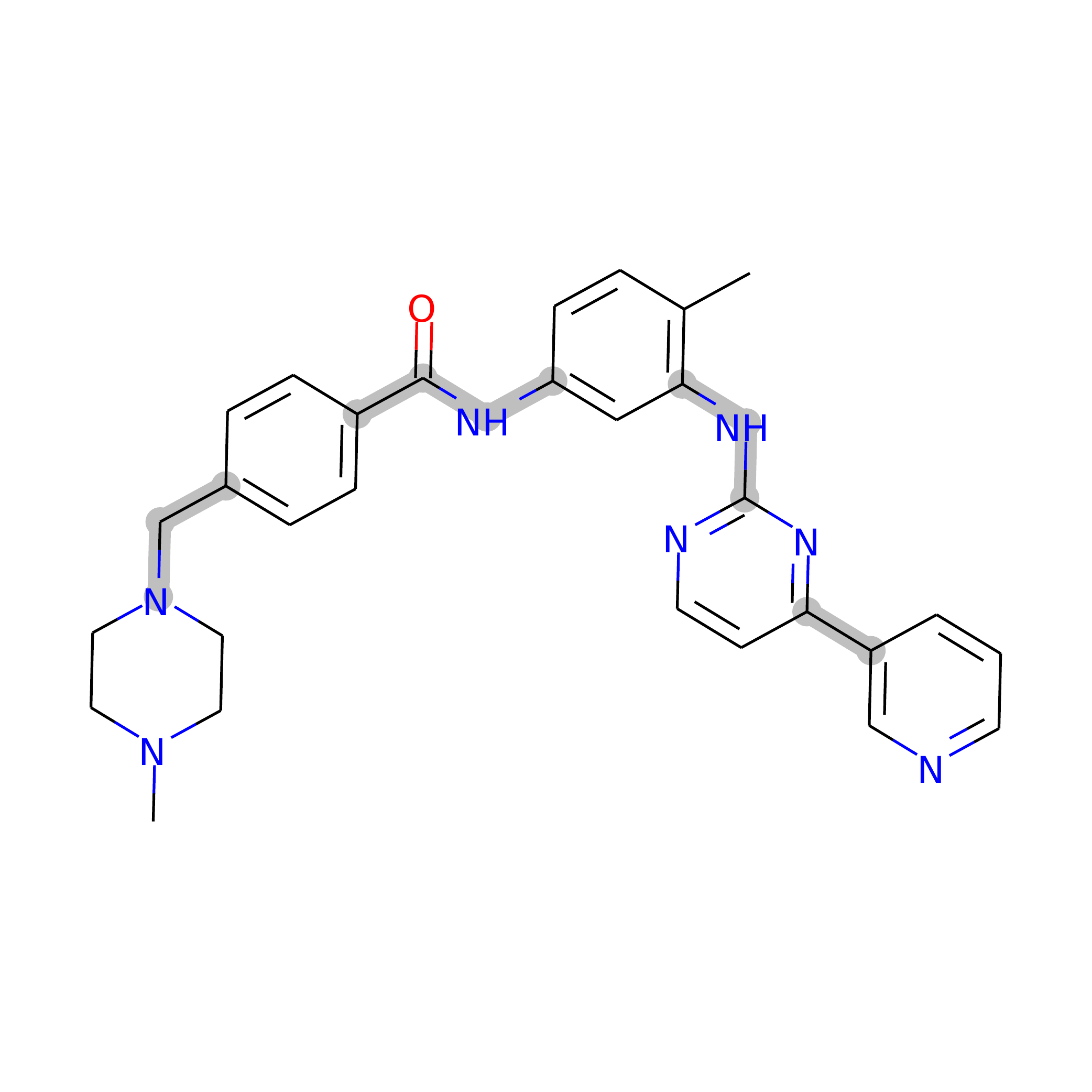}
\caption{Imatinib (STI)}
\end{subfigure}
\begin{subfigure}[b]{0.41\textwidth}
\includegraphics[width=\textwidth]{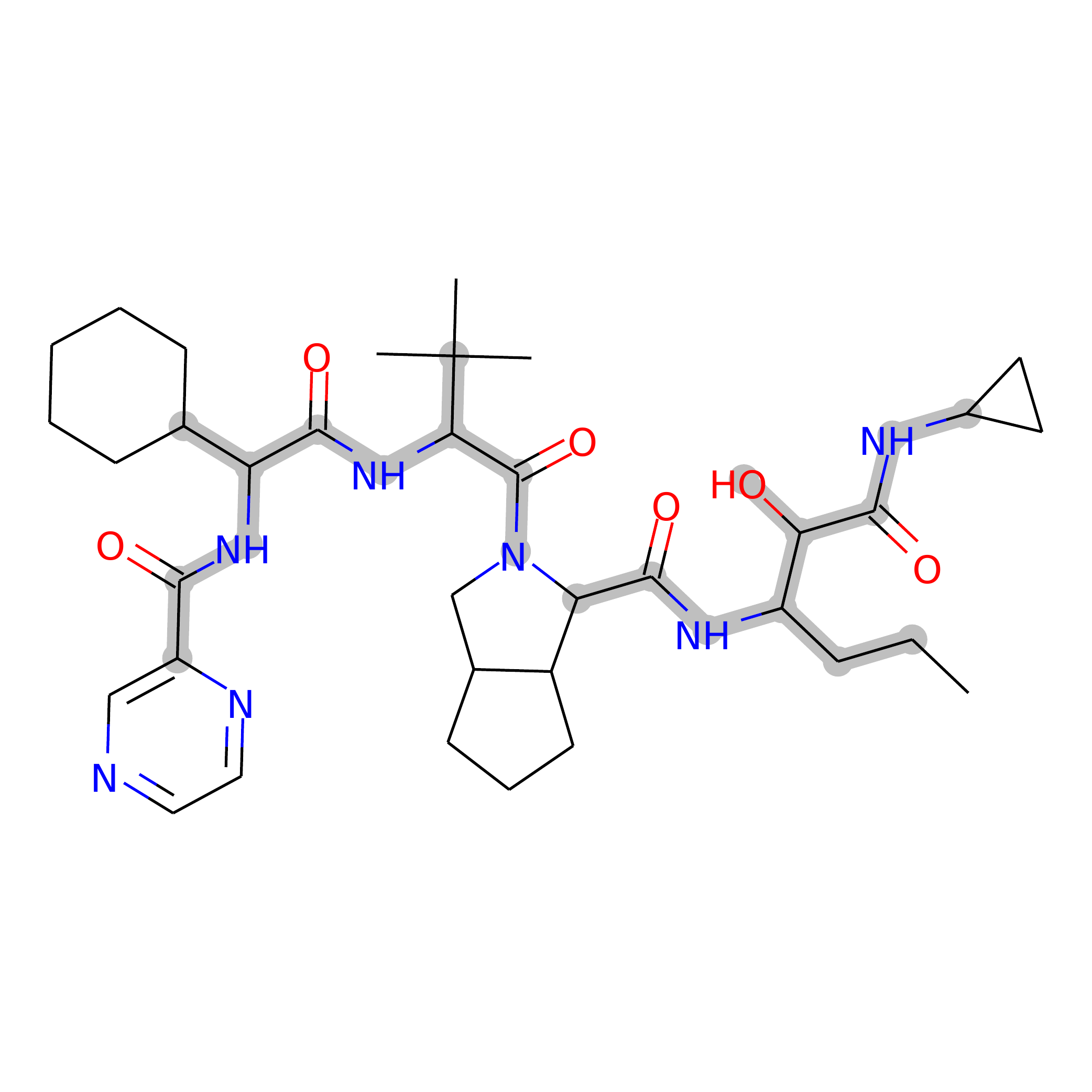}
\caption{Telaprevir (SV6)}
\end{subfigure}
\caption{(a)--(d) structures of TopDrugs. The parameterized dihedral angles are shown in grey.}
\label{fig:TopDrugs}
\end{figure}

The second dataset, called TopDrugs, consists of four molecules (Table~\ref{tab:TopDrugs} and Figure~\ref{fig:TopDrugs}) selected to represent realistic application scenarios in computational \gls{sbdd}. For that, we looked at recent top-selling drugs\cite{TopDrugsWiki, TopDrugsIgea}, and selected them based on four criteria: small size (<110~atoms), contain only H, C, N, and O elements, have rotable bonds, and present in the Protein Data Bank (i.e. a reasonable structure is available).

\begin{table}
\begin{threeparttable}
\caption{Overview of TopDrugs}
\label{tab:TopDrugs}
\begin{tabular}{llrr}
\toprule
Name         & PDB chemical ID  & Number of atoms & Number of dihedral angles\textsuperscript{$\dagger{}$} \\ \midrule
Lenalidomide & LVY              & 32              & 2                         \\
Apixaban     & GG2              & 59              & 6                         \\
Imatinib     & STI              & 68              & 8                         \\
Telaprevir   & SV6              & 104             & 20                        \\ \bottomrule
\end{tabular}
\begin{tablenotes}
\item $\dagger{}$ as described in section \refbyname{sec:diheds}
\end{tablenotes}
\end{threeparttable}
\end{table}

The last dataset, called ZINC, consists of 1000 molecules from ZINC12 database, containing over 20 million commercially-available organic molecules\cite{zinc12}. The molecules were selected with a three-step procedure: first, molecules containing only H, C, N, and O elements were selected; second, they were grouped according to the number of parameterizable dihedral angles (as described in section \refbyname{sec:diheds}); finally, 100 molecules were selected randomly from each group (from 1 to 10 dihedral angles). The complete list of the molecules is available in SI (Table~S1--S2).

\section{Results}

\emph{Parameterize} is evaluated in three aspects. First, we measure the accuracy of \gls{nnp} energies. Then, we assess the overall quality of \gls{ff} parameters. Finally, we check parameterization time, reliability, and scaling.

\subsection{Accuracy of \gls{nnp}}

The energies of \gls{nnp}, GAFF, and GAFF2 are compared with the \refenergyshort~ results using the rotamers of \citet{Sellers} and TopDrugs (Table~\ref{tab:TopDrugs} and Figure~\ref{fig:TopDrugs}) molecules. The dihedral angles were selected and rotamers were generated as described in section \refbyname{sec:diheds}. All energy profiles are available in SI (Figure~S1--S49).

\begin{figure*}
\centering
\includegraphics[width=.6\textwidth]{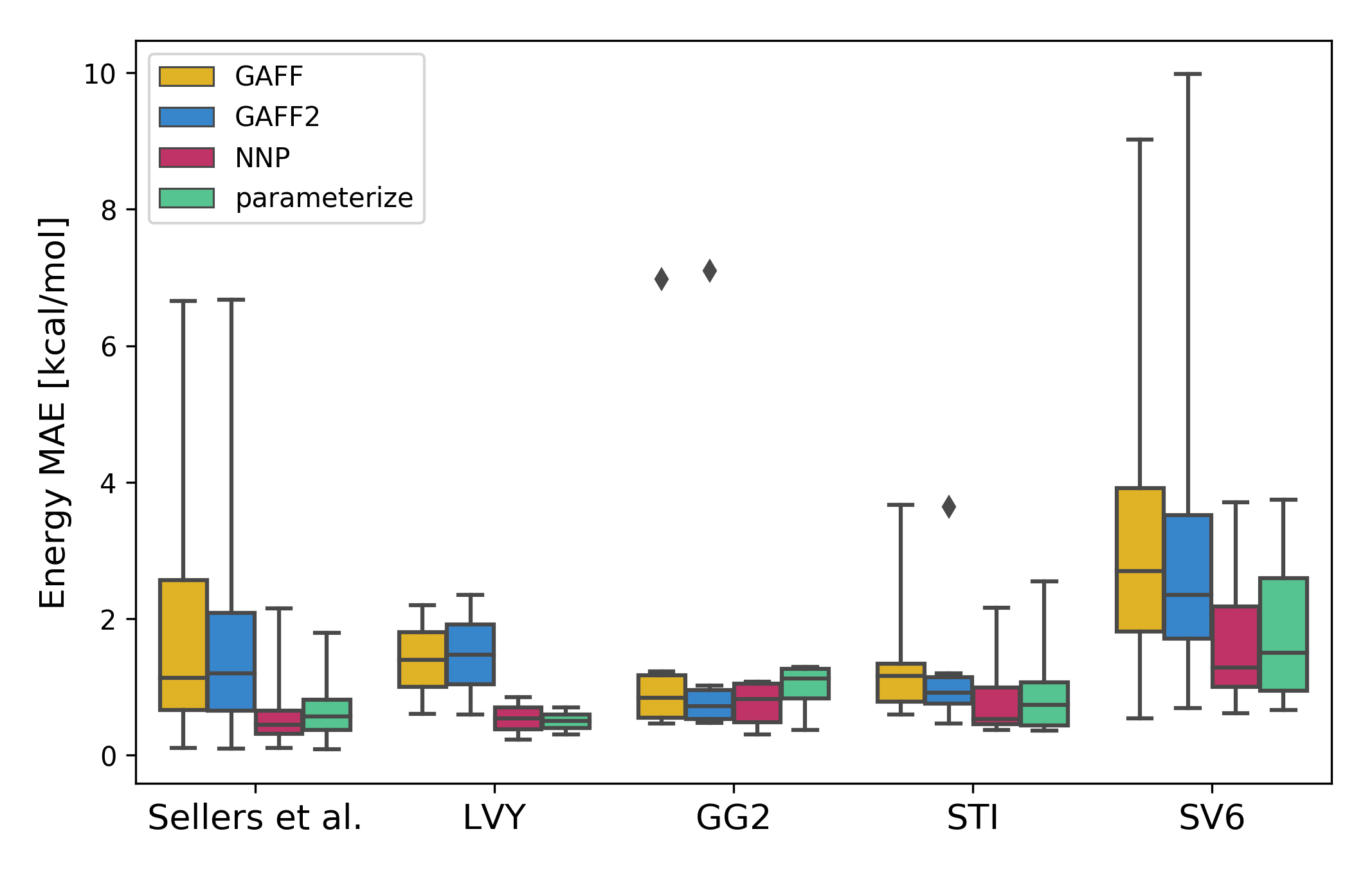}
\caption{Comparison of GAFF, GAFF2, \gls{nnp}, and \emph{Parameterize} energies with respect to the \refenergyshort~results using the rotamers of \citet{Sellers} and TopDrugs (Table~\ref{tab:TopDrugs} and Figure~\ref{fig:TopDrugs}) molecules.}
\label{fig:energy_comparison}
\end{figure*}

\begin{figure*}
\centering
\includegraphics[width=.6\textwidth]{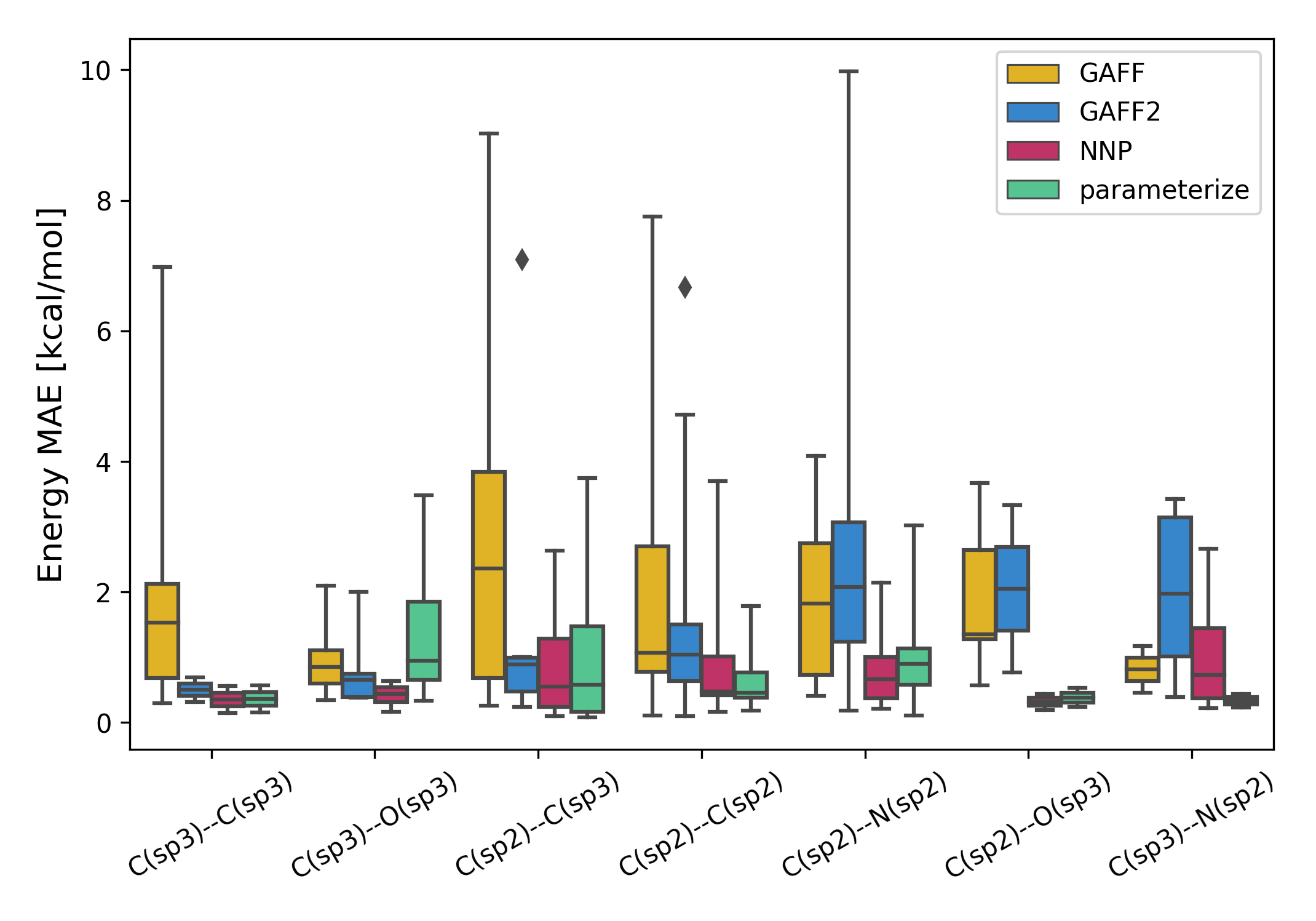}
\caption{Comparison of dihedral angle types in terms of GAFF, GAFF2, \gls{nnp}, and \emph{Parameterize} energies with respect to the \refenergyshort~results. The dihedral angles are typed according to the central atom elements and hybridizations.}
\label{fig:dihedral_comparison}
\end{figure*}

\gls{nnp} is more accurate than GAFF and GAFF2 (Figure~\ref{fig:energy_comparison}). In the case of \citet{Sellers} dataset, the \gls{mae} of energy is \SELLERSNNP~kcal/mol, while for GAFF and GAFF2 it is \SELLERSGAFF~ and \SELLERSGAFFTWO~kcal/mol, respectively. In the case of TopDrugs, the trend is the same, but the largest molecule (SV6) has larger errors: \SVSIXNNP~kcal/mol for \gls{nnp}, \SVSIXGAFF~kcal/mol for GAFF, and \SVSIXGAFFTWO~kcal/mol for GAFF2.

In terms of the dihedral angle types (Figure~\ref{fig:dihedral_comparison}), \gls{nnp} is more accurate than GAFF and GAFF2 in almost all cases. Noticeably, the accuracy of \gls{nnp} is higher for C(sp\textsuperscript{3})--C(cp\textsuperscript{3}), C(sp\textsuperscript{3})--O(cp\textsuperscript{3}), and C(sp\textsuperscript{2})--O(cp\textsuperscript{3}); than C(sp\textsuperscript{2})--C(cp\textsuperscript{3}), C(sp\textsuperscript{2})--C(cp\textsuperscript{2}), C(sp\textsuperscript{2})--N(cp\textsuperscript{2}), and C(sp\textsuperscript{3})--N(cp\textsuperscript{2}) dihedral angles. The latter dihedral angles (containing atoms with  the sp\textsuperscript{2} hybridization) represent more diverse chemical groups (e.g. aromatic and conjugated systems).

\subsection{Quality of \gls{ff} parameters}

All molecules from \citet{Sellers} and TopDrugs (Table~\ref{tab:TopDrugs} and Figure~\ref{fig:TopDrugs}) datasets were parameterized with our method and the energies of theirs rotamers were computed with the new \gls{ff} parameters. All energy profiles are available in SI (Figure~S1--S49).

The accuracy of the new \gls{ff} parameter is close to the \gls{nnp} results (Figure~\ref{fig:energy_comparison}). In the case of \citet{Sellers} dataset, the \gls{mae} with respect to the \refenergyshort~ results is \SELLERSPARAMETERIZE~kcal/mol, which is comparable with the \gls{mae} of \gls{nnp} itself (\SELLERSNNP~kcal/mol). For TopDrugs molecules, the trend is the same: the new \gls{ff} parameters reproduce the \gls{nnp} results closely and are more accurate than GAFF2 for all the molecules except GG2. Also, for different dihedral angles types, the \gls{mae} does not exceed 0.5~kcal/mol (Figure~\ref{fig:dihedral_comparison}).

\begin{figure}
\centering
\begin{subfigure}[b]{0.4\textwidth}
\includegraphics[width=\textwidth]{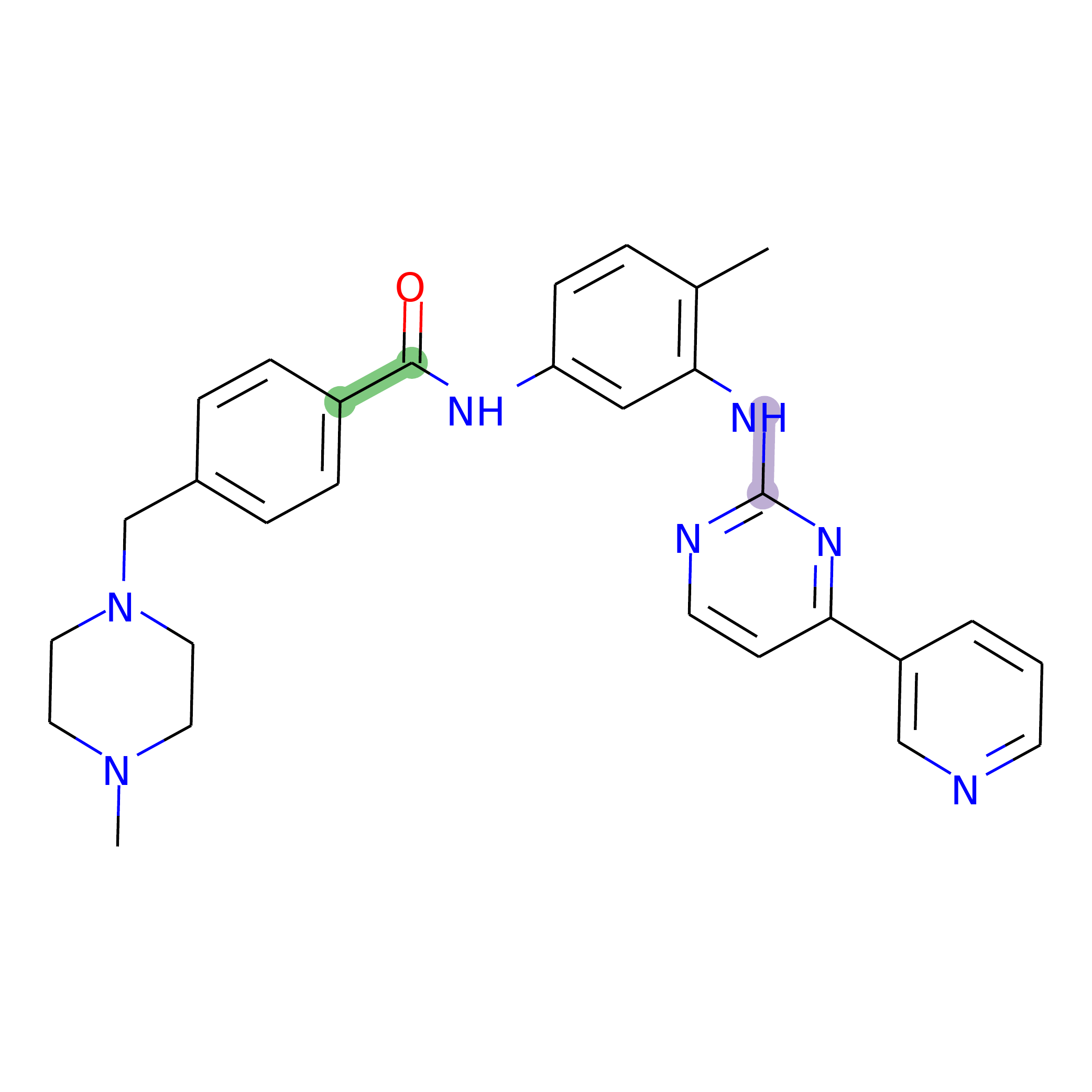}
\caption{}
\end{subfigure}\\
\begin{subfigure}[b]{0.45\textwidth}
\includegraphics[width=\textwidth]{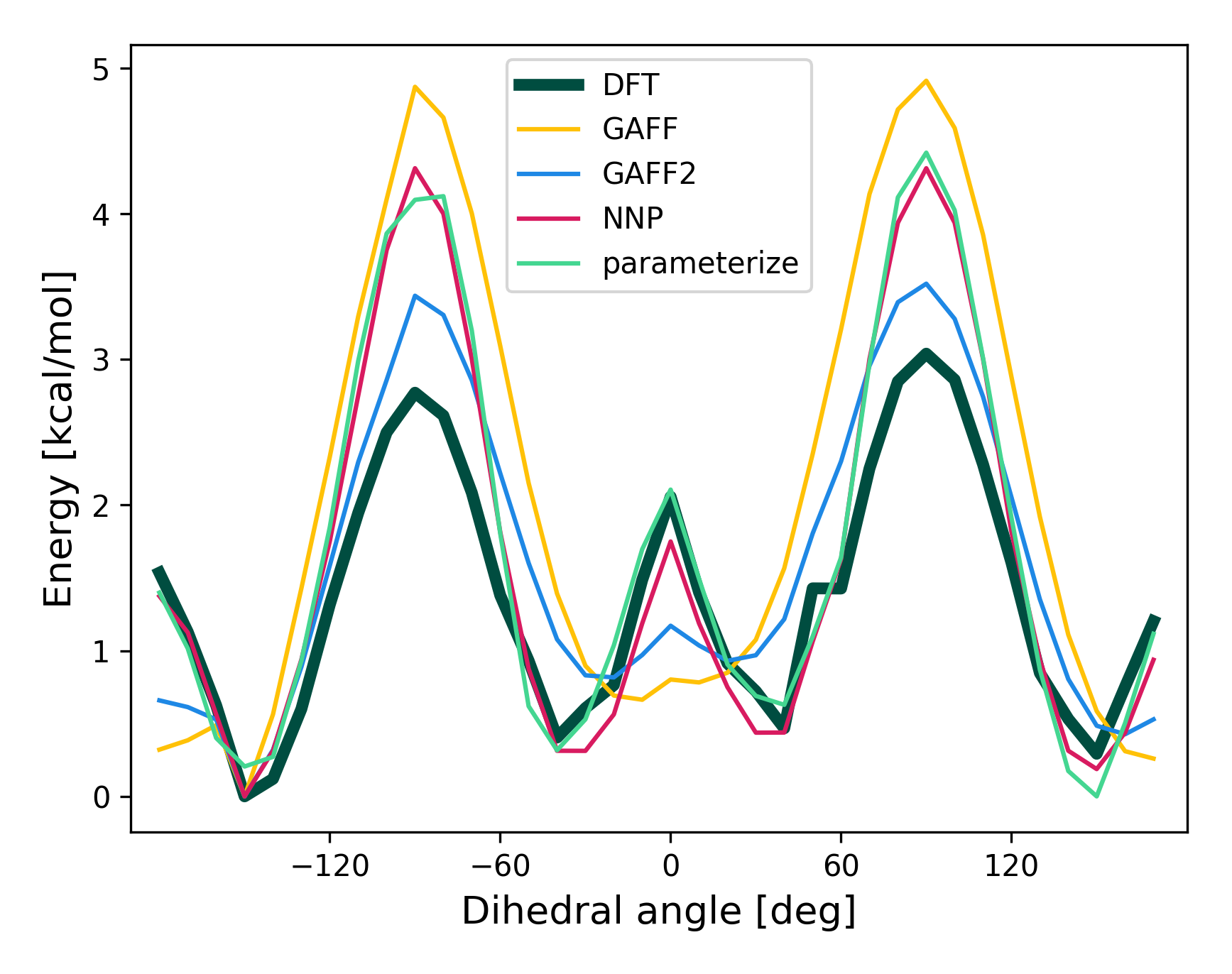}
\caption{}
\end{subfigure}\\
\begin{subfigure}[b]{0.45\textwidth}
\includegraphics[width=\textwidth]{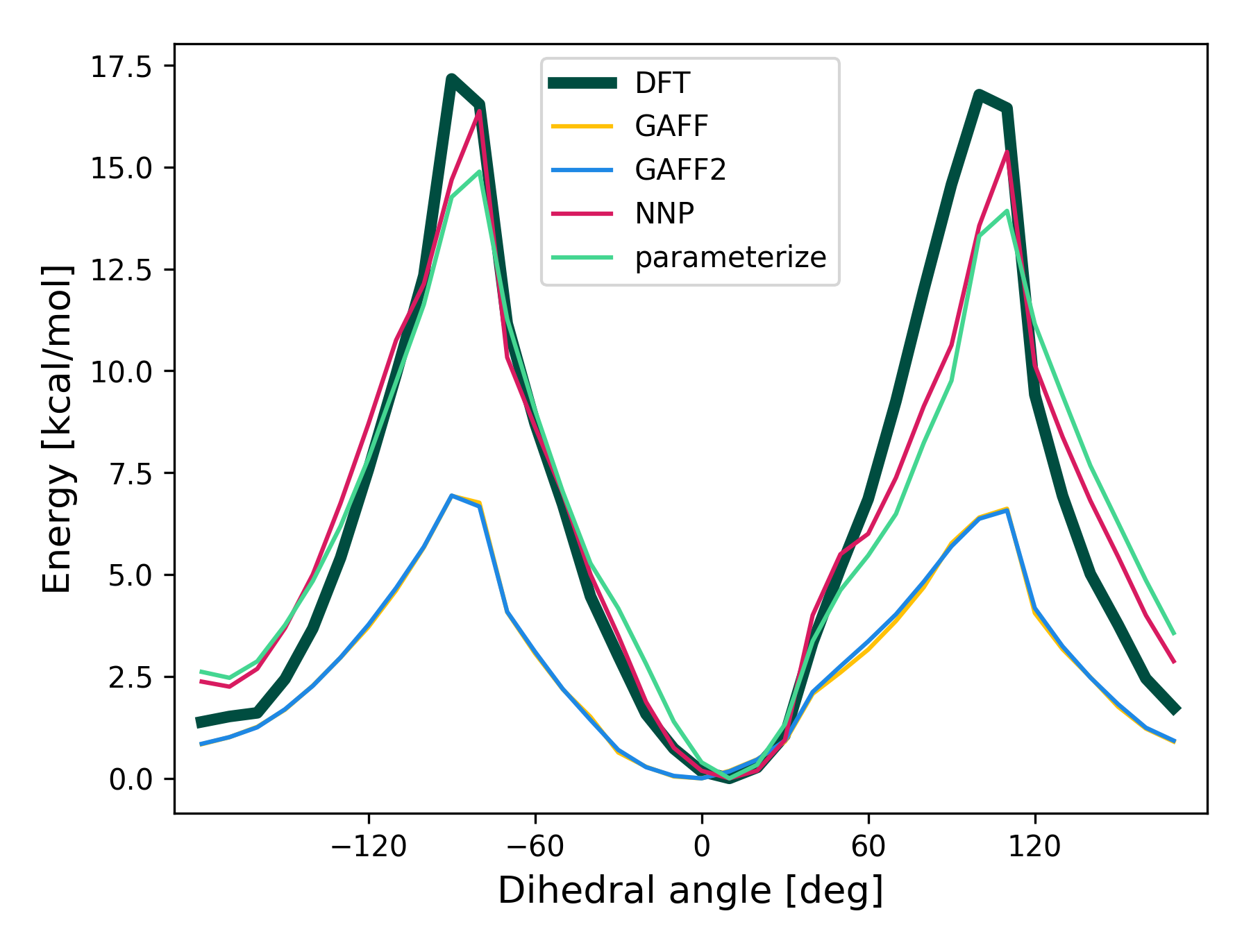}
\caption{}
\end{subfigure}
\caption{Comparison of GAFF, GAFF2, \gls{nnp}, and \emph{Parameterize} energy profiles with respect to the \refenergyshort~ results using two selected dihedral angle of STI (a). The dihedral angles are shown in green (b) and purple (c).}
\label{fig:selected_profiles}
\end{figure}

Furthermore, we inspect the energy profiles of a few selected dihedral angles of STI (Figure~\ref{fig:selected_profiles}). The dihedral angles were selected to illustrate the cases where the new \gls{ff} parameters can reproduce the results of \refenergyshort{} and where they cannot. Note that the energy profiles include all the energy terms and not just the dihedral angle terms.

Finally, the overall quality of \gls{ff} parameters (not just dihedral angle parameters) is checked by minimizing the TopDrugs molecules. The minimization is started and RMSD is computed with respect to the \refenergyshort-minimized structure. None of the molecules undergoes any significant conformation changes and RMSDs are 0.4~\si{\angstrom} or less (Table~\ref{tab:rmsd-from-qm}), except SV6.

\begin{table}
\centering
\caption{Comparison of the minimized TopDrugs molecules. The RMSDs are computed with respect to the \refenergyshort-minimized structures.}
\label{tab:rmsd-from-qm}
\begin{threeparttable}
\begin{tabular}{lccc}
\toprule
\multicolumn{1}{c}{\multirow{2}{*}{Molecule}} & \multicolumn{3}{c}{RMSD [\si{\angstrom}]} \\ \cline{2-4} 
\multicolumn{1}{c}{}                        & \multicolumn{1}{l}{GAFF} & \multicolumn{1}{l}{GAFF2} & \multicolumn{1}{l}{\emph{Parameterize}} \\
\midrule
LVY                                & 0.305                     & 0.301                     & 0.190                             \\
GG2                                    & 0.292                     & 0.294                     & 0.376                             \\
STI                                    & 0.156                     & 0.156                     & 0.126                             \\
SV6                                  & 0.641                      & 0.620                       & 0.827                              \\
\bottomrule
\end{tabular}
\end{threeparttable}
\end{table}

\subsection{Parameterization time and reliability}


The benchmark of \emph{Parameterize} is performed with the example molecule from Figure~\ref{fig:demo_molecule} (18~atoms and two parameterizable dihedral angles) on \textbf{a single core} of a 2.1~GHz processor (Intel Xeon E5-2620). Note that no \gls{gpu} is used.

The total parameterization time is 31~s (Table~\ref{tab:speed}), where the energy calculations with \gls{nnp} take just 3~s. For comparison, the same calculations with \refenergyshort{} take $\sim$0.7~min per rotamer. Thus, the parameterization time would be $\sim$50~min ($2 \times 36$~rotamers)!

\begin{table}
\centering
\caption{Parameterization time (in seconds) for the molecule from Figure~\ref{fig:demo_molecule} with different reference energy methods (\refenergyshort~ and \gls{nnp}).}
\label{tab:speed}
\begin{threeparttable}
\begin{tabular}{lcc}
\toprule
\multicolumn{1}{c}{\multirow{2}{*}{Procedure}} & \multicolumn{2}{c}{Reference} \\
\cline{2-3}
\multicolumn{1}{c}{}                  & \refenergyshort & \gls{nnp} \\
\midrule
GAFF2 parameters   &    1 &  1\\
AM1-BCC charges    &    2 &  2\\
Dihedral scans     &   13 & 13\\
Reference energies & 3024 &  3\\
Parameter fitting  &    7 &  7\\
Other              &    5 &  5\\
\midrule
Total              & 3052 & 31\\
\bottomrule
\end{tabular}
\end{threeparttable}
\end{table}


The reliability of \emph{Parameterize} is checked with the ZINC molecules\cite{zinc12}. Out of 1000 molecules, 949 molecules completed successful, 14 molecules had large errors (i.e. the \gls{mae} of dihedral parameter fitting was >2.0~kcal/mol), and 37 molecules failed (Table S2). The failures occurred in the dihedral angle parameter fitting (26 molecules), the rotamer generation (7 molecules), and the initial atom types and \gls{ff} parameters assignment (4 molecules). Note that due to limited computational resources, we do not compare with \refenergyshort{} results.

\begin{figure}
\centering
    \begin{subfigure}{0.45\textwidth}
        \includegraphics[width=\textwidth]{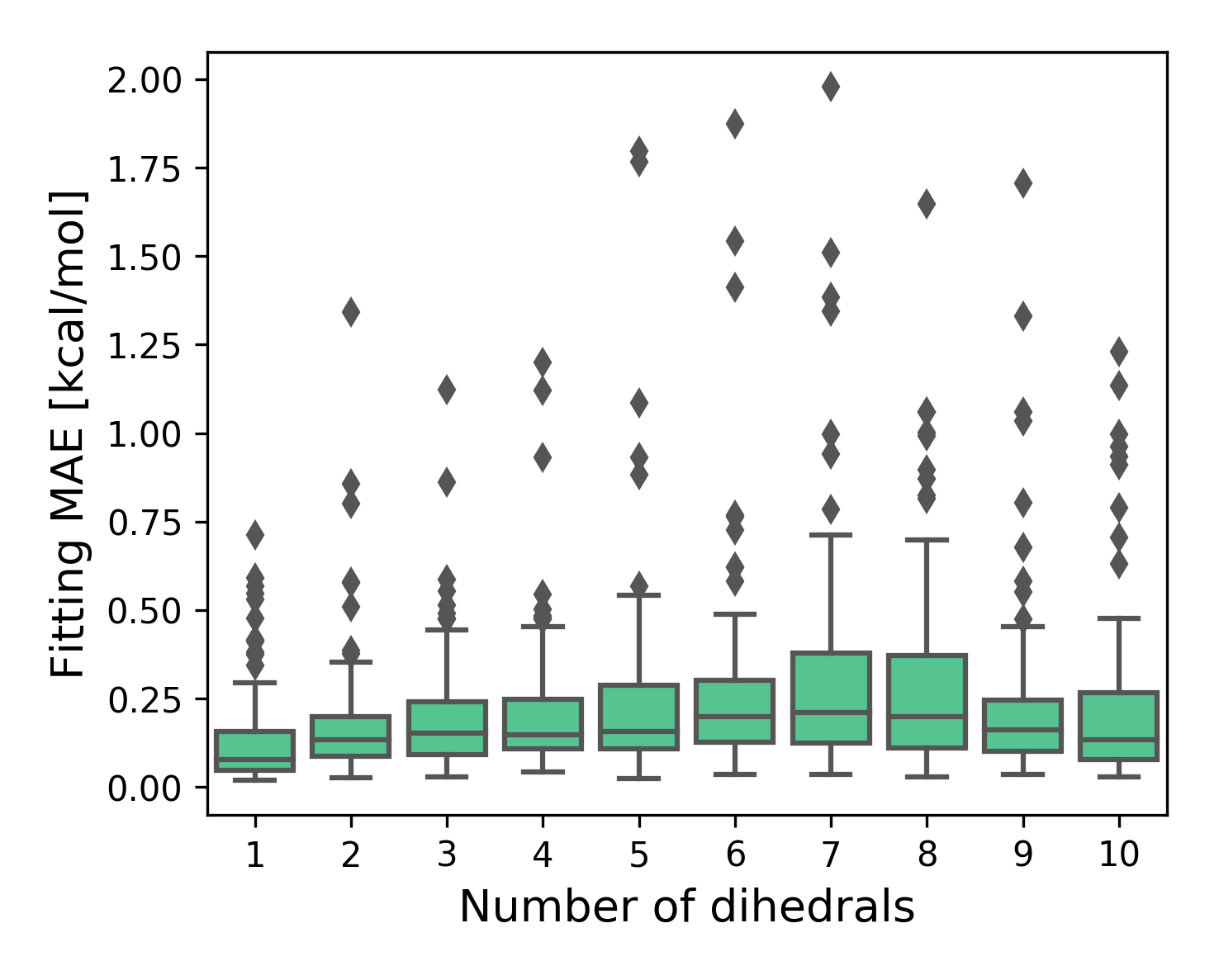}
        \caption{}
        \label{fig:zinc_errors}
    \end{subfigure}
    \begin{subfigure}{0.45\textwidth}
        \includegraphics[width=\textwidth]{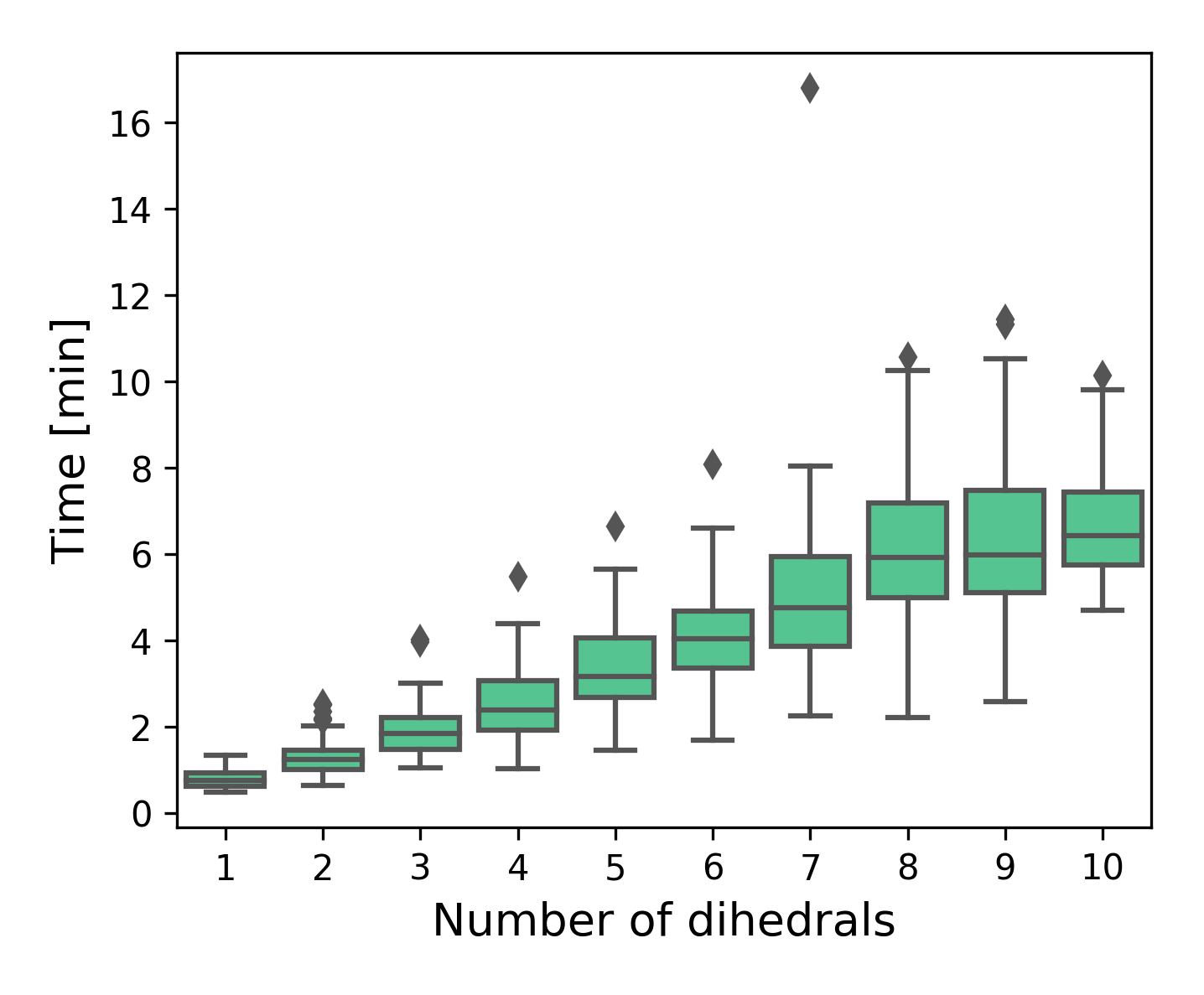}
        \caption{}
        \label{fig:zinc_times}
    \end{subfigure}
\caption{\emph{Parameterize} scaling with respect to the number of dihedral angles in term of fitting \gls{mae} (a) and parameterization time (b). The statistics are computed from the results of the ZINC molecules.}
\label{fig:zinc_results}
\end{figure}

Finally, the scaling of \emph{Parameterize} is measured with the groups of ZINC molecules containing different numbers (from 1 to 10) of parameterizable dihedral angles. The overall \gls{mae} of the dihedral angle parameter fitting is \ZINCMAE~kcal/mol and it does not depend significantly on the number of dihedral angles (Figure~\ref{fig:zinc_errors}). Meanwhile, the parameterization time grows linearly with the number of dihedral angles (Figure~\ref{fig:zinc_times}). The results of all the ZINC molecules are available in SI (Table~S1--S2).

\section{Discussion}


We have demonstrated that \glspl{nnp} can achieve better accuracy than \gls{mm} with GAFF2\cite{Wang2004} parameters (Figure~\ref{fig:energy_comparison} and~\ref{fig:dihedral_comparison}). A question is why \glspl{nnp} cannot be used directly in \gls{md}. Here, we argue that, despite the progress in developing \glspl{nnp}, they are not yet ready to replace \gls{mm}.
First, the BP symmetry functions used in ANI-1x are good at describing the chemical environment at short range (<\SI{6}{\angstrom}), but they intrinsically lack long-range interactions\cite{ani1xcc, smith_less_2018}. It can be seen that the results of the largest molecule (SV6) are the worst (Figure~\ref{fig:energy_comparison} and Table~\ref{tab:rmsd-from-qm}). Also, the training of \glspl{nnp} with larger molecules is problematic due to rapidly growing computation cost of \gls{qm}, which is needed to generate training data. However, the new types of chemical environment featurization, introduced by SchNet\cite{SchNet} and AIMNet~\cite{aimnet}, may improve the long-range interactions.
Second, current \glspl{nnp} support limited number of elements. For example, ANI-1x\cite{ani1xcc, smith_less_2018} and AIMNet~\cite{aimnet} implementations only support four (H, C, N, and O) and six (H, C, N, O, F, and S) elements, respectively. That is not sufficient for typical \gls{sbdd} simulations, which may contain H, C, N, O, P, S, halogens, and metal cations.
Thus, \emph{Parameterize} bridges the gap between \glspl{nnp} and \gls{md}. It leverages the accuracy and speed of \glspl{nnp} to improve the \gls{ff} parameters of individual molecules, while benefiting from remaining compatible with existing \gls{md} software and established \glspl{ff}. Moreover, \emph{Parameterize} allows a quick adaptation of improved \gls{nnp} methods, when they will become available, or Psi4\cite{psi4} can be used directly for \gls{qm} energies, if the computational cost is acceptable.


Further, we have demonstrated that the accuracy of the \gls{ff} for drug-like molecules can be improved just fitting dihedral angles parameters (Figure~\ref{fig:energy_comparison} and~\ref{fig:selected_profiles}), while the rest parameters are taken from a general \gls{ff} (GAFF2 in our case). This simplifies the parameterization procedure and allows a quick adaptation to other \gls{ff} families. For example, \emph{Parameterize} could be adapted to CHARMM~\cite{mackerell1998all, huang2013charmm36} using CGenFF~\cite{vanommeslaeghe2010charmm, charmm_note}.


In some cases, the new \gls{ff} parameters cannot reproduce the \refenergyshort~ results (Figure~\ref{fig:selected_profiles}). A part of the errors can be attributed to the accuracy of \glspl{nnp}, but, in addition, we have identified at least three other components of the parameterization procedure (Figure~\ref{fig:param_graph}), which are going to be improved in the future.
The atomic charges are assigned with AM1-BCC\cite{bcc_charges_1, bcc_charges_2}. It would be more accurate to use the RESP\cite{esp_charges, resp_charges} charges, but it requires expensive \gls{qm} calculation. This could be solved with \gls{ml}-based charge assignment methods\cite{ml-partial-charges, transferable-charge-assignment}. Also, the atomic charges are conformation dependent, so it is better to average them over a representative ensemble of conformations rather than just use a single conformation.
In addition, the more accurate Lennard-Jones parameters by \citet{boulanger2018optimized} could be used, rather than the GAFF2 parameters\cite{Wang2004}.
Moreover, we have found that, for the dihedral angle scanning and the following parameter fitting, it is better to use \gls{mm}-optimized conformations than more accurate \gls{qm}-minimized ones. Partially, this can be solved by fitting more soft mode parameters (e.g. improper dihedral angles), but ultimately the accuracy is limited by the potential forms of \gls{ff}. For example, AMBER~\cite{cornell1995second, maier2015ff14sb} family \glspl{ff} do not include explicit potentials to model correlation between dihedral angles, hydrogen bonds, $\pi$--$\pi$ stacking, etc.


Finally, the overall reliability of \emph{Parameterize} is $\sim$95\%. The majority of failures occur in the fitting of the dihedral angles parameters, where the optimizer either fails to find good quality parameters or fails completely. Another problem is that the parameter space is non-linear and multi-modal, which requires a global optimizer. Currently, it is the main bottleneck limiting the speed of \emph{Parameterize} (Table~\ref{tab:speed}). Our proposed optimization algorithm is an effort to balance reliability and scaling with the size of molecules (Figure~\ref{fig:zinc_results}). However, for the larger molecules (e.g. biopharmaceuticals containing hundreds of atoms and tens of dihedral angles), we may need a different approach, such as an automated molecule fragmentation scheme.

\section{Conclusion}


We have developed an \gls{ff} parameterization procedure which can utilize either \gls{qm} calculations or \gls{nnp} energy predictions. The quality of the Parameterize method using \glspl{nnp} is evaluated using 45 druglike fragments by \citet{Sellers} and four real drug molecules. Our validation using ANI-1x\cite{ani1xcc, smith_less_2018} as a \gls{nnp} is promising, achieving better accuracy than GAFF2\cite{Wang2004} as the \gls{mae} is reduced from \SELLERSGAFFTWO~kcal/mol to \SELLERSNNP~ kcal/mol in the case of \citet{Sellers} fragments. Furthermore, the accuracy of the \gls{ff} is improved by fitting selected dihedral angle parameters to reproduce \gls{nnp} energies (\gls{mae} is \SELLERSPARAMETERIZE~kcal/mol in the case of \citet{Sellers} fragments), while the rest of the parameters are taken from GAFF2. Finally, it demonstrates that small molecules parameters can be obtained in minutes (rather than hours, if \gls{dft} had been used) with $\sim$95\% reliability for the randomly selected ZINC12\cite{zinc12} molecules.

The current version of \emph{Parameterize} is available on the drug-discovery platform \emph{PlayMolecule} (\url{www.playmolecule.org}). The current limitation of atom elements of \glspl{nnp} like ANI-1x limits its applicability to benchmark molecules. Yet, once this problem is solved, Parameterize using \glspl{nnp} would fill the gap between the general \gls{ff} (e.g. GAFF~\cite{Wang2004} and CGenFF~\cite{vanommeslaeghe2010charmm}), which can generate parameters in seconds but with limited accuracy and \gls{qm}-based methods (e.g. GAAMP~\cite{Huang2013}, ffTK~\cite{Mayne2013}, and ATB~\cite{Malde2011}), which provide more accurate parameters but require hours or even days of computation time.

\begin{acknowledgement}
The authors thank Alberto Cuzzolin, Toni Giorgino, Dominik Lemm, Gerard Mart\'{i}nez-Rosell, Davide Sabbadin and Olexandr Isayev for helpful discussions.
This project has received funding from the European Union's Horizon 2020 research and innovation programme under grant agreements No~739649 (Cloud-HTMD project) and No~675451 (CompBioMed project).
\end{acknowledgement}

\section{Conflict of Interest Disclosure}
All the authors are employed by Acellera.

\begin{suppinfo}
\begin{itemize}
  \item Figure S1--S49: the structures of the molecules from \citet{Sellers} and TopDrugs datasets and their energy profiles (DFT, NNP, GAFF, GAFF2, \emph{Parameterize}) of parameterized dihedral angles.
\item Table S1--S2: the selected molecules from the ZINC dataset, their parameterization results, and failures.
\end{itemize}
\end{suppinfo}

\bibliography{references}

\begin{mcitethebibliography}{49}
\bibitem[res()]{restriction}
Due to commercial interests, the free public access is restricted to the
  molecules up to 50 atoms and 8parameterizable dihedral angles.\relax
\mciteBstWouldAddEndPunctfalse
\mciteSetBstMidEndSepPunct{\mcitedefaultmidpunct}
{}{\mcitedefaultseppunct}\relax
\EndOfBibitem
\end{mcitethebibliography}

\end{document}